\documentclass[12pt]{article}

\usepackage{float}
\usepackage[round,comma,authoryear]{natbib}
\usepackage{amsmath, amssymb, amsfonts}
\RequirePackage{amsthm}
\usepackage{algorithm}
\usepackage{algorithmic}
\usepackage{hyperref}
\usepackage{amsmath, amssymb, amsfonts} 
\usepackage{graphicx, subfigure} 
\usepackage{booktabs}            
\usepackage{multirow}            
\usepackage{color,soul}
\usepackage{mdframed}
\usepackage{setspace}
\usepackage[utf8]{inputenc}
\usepackage{caption}
\def\bSig\mathbf{\Sigma}

\newcommand{\bigCI}{\mathrel{\text{\scalebox{1.07}{$\perp\mkern-10mu\perp$}}}}
\graphicspath{{./figs/}}

\renewcommand\appendix{\par
  \setcounter{section}{0}
  \setcounter{subsection}{0}
  \setcounter{figure}{0}
  \setcounter{table}{0}
  \renewcommand\thesection{Appendix \Alph{section}}
  \renewcommand\thefigure{\Alph{section}\arabic{figure}}
  \renewcommand\thetable{\Alph{section}\arabic{table}}
}

\usepackage[utf8]{inputenc}
\usepackage[english]{babel}
\newtheorem{theorem}{Theorem}[section]
\newtheorem{lemma}[theorem]{Lemma}

\title{Dynamic Chain Graph Models for Ordinal Time Series Data}

\author{P. Behrouzi\\
  Wageningen University and Research Centre\\
  \texttt{pariya.behrouzi@wur.nl}
    \and F. Abegaz\\
    University of Liège\\
    \texttt{Y.FAbegaz@ulg.ac.be}\\
  \and E. C. Wit\\
  University of Groningen\\
  \texttt{e.c.wit@rug.nl}
  }

\date{}
\begin{document}

\maketitle

\begin{abstract}
This paper introduces sparse dynamic chain graph models for network inference in high dimensional non-Gaussian time series data. The proposed method parametrized by a precision matrix that encodes the intra time-slice conditional independences among variables at a fixed time point, and an autoregressive coefficient that contains dynamic conditional independences interactions among time series components across consecutive time steps. The proposed model is a Gaussian copula vector autoregressive model, which is used to model sparse interactions in a high-dimensional setting. Estimation is achieved via a penalized EM algorithm. In this paper, we use an efficient coordinate descent algorithm to optimize the penalized log-likelihood with the smoothly clipped absolute deviation penalty. We demonstrate our approach on simulated and genomic datasets. The method is implemented in an {\tt R} package \textbf{tsnetwork}.

\textbf{Key words:} Chain graph models; time-series data; Latent variable; Gaussian Copula; SCAD penalty ; $L_1$ penalty; penalized likelihood; Vector autoregressive model.
 \end{abstract}

\section{Introduction}
\label{into}
Graphical models are an efficient tool for modeling and inference in high dimensional settings. Directed acyclic graph (DAG) models, known as Bayesian networks \citep{lauritzen1996graphical}, are often used to model asymmetric cause-effect relationships. Models represented by undirected graphs are used to model symmetric relationships, for instance gene regulatory networks. 

Some graphical models are able to represent both asymmetric and symmetric relationships simultaneously. One such model so-called chain graph model \citep{lauritzen1996graphical, lauritzen1989graphical} which is a generalization of directed and undirected graphical models. Chain graph models contain a mixed set of directed and undirected edges. The vertex set of a chain graph can be partitioned into chain components where edges within a chain component are undirected whereas the edges between two chain components are directed and point in the same direction. Recently, chain graph models are considered in a time series setting \citep{abegaz2013sparse, gao2010latent, dahlhaus2003causality}. 

There is a rich literature on reconstructing undirected graph for continuous data, categorical data, and mixed categorical and continuous data \citep{Behrouzi2017detecting, mohammadi2015bayesian, dobra2011copula, hoff2007extending} and similarly for directed acyclic graphs \citep{colombo2012learning, kalisch2007estimating}. Recently, \cite{abegaz2013sparse} have proposed a method based on chain graph model for analyzing time course continuous data, like gene expression data. However, many real-world time series data are not continuous, but are categorical or mixed categorical and continuous. Until now constructing dynamic networks for non-continuous time series data has remained unexplored. Here, we develop a method to explore dynamic or delayed interactions and contemporaneous interactions for time series of categorical data and time series of mixed categorical and continuous data.

The proposed method is based on chain graph models, where the ordered time steps build a DAG of blocks and each block contains an undirected network of variables under consideration at that time point. The method developed in this paper is designed to analyze the nature of interactions present in repeated multivariate time series mixed categorical and continuous data, where we use time series chain graphical models to study the conditional independence relationships among variables at a fixed time point and “causal” relationship among time series components across consecutive time steps. The concept of causality that we use is the concept of Granger causality \citep{granger1969investigating}, which exploits the natural time ordering to achieve a “causal” ordering of the variables in multivariate time series. The idea of this causality concept is based on predictability, where one time series is said to be Granger causal for another series if the latter series to be better predicted using all available information than if the information apart from the former series had been used. Our inference procedure not only enforces sparsity on interactions within each time step, but it also between time steps; this feature is particularly realistic in a real-world dynamic networks setting. 

We proceed as follow: in section \ref{method}, we explain the method where we first introduce dynamic chain graph models in section \ref{DynamicCGM}, then we propose the Gaussian copula for mixed scale time series data in section \ref{GCGM}. In sections \ref{likelihood} and \ref{Inference} we define a model for underlying multivariate time series components and we explain the procedure of penalized inference based on the L1 norm and smoothly clipped absolute deviation (SCAD) penalty terms. In section \ref{modelSelection} we present a method for obtaining the log-likelihood of the observed mixed scale time series component under the penalized EM algorithm and we proceed with model selection for tuning the penalty terms. In section \ref{Sim} we study the performance of the proposed dynamic chain graph model under different scenarios. Furthermore, we compare its performance with the other available methods. The proposed method is demonstrated in section \ref{application} to investigate the course of depression and anxiety disorders.
\section{Methods}
\label{method}
\subsection{Dynamic chain graph models}
\label{DynamicCGM}
A chain graph is defined as $G = (V, E)$ where $V$ is a set of vertices (nodes) and $E$ is a set of ordered and unordered pairs of nodes, called edges, which contains the directed and undirected interactions between pairs of nodes. A dynamic chain graph model is associated with a time series chain graph model, where the dependence structure of the time series components can be divided into two sets: \emph{intra time-slice dependencies}, which are represented by undirected edges that specify the association among variables in a fixed time step, and a set of \emph{inter time-slice dependencies}, which are represented by associations among variables across consecutive time steps. Links across time steps are directed pointing from a set of nodes at a previous time step, $V_{(t-1)}$, to nodes at the current time step, $V_t$. The dynamic chain graph models in our modeling framework relates the time series components at time $t$ to only that of at time $t - 1$, but this can be easily extended to a higher order ($d \ge 2$) time steps. 

Let $\mathbf{Y}(t) = (Y_1(t), \ldots, Y_p(t) \acute{)}, t=1,\ldots, T$ be an p-dimensional time series vector representation of $p$ variables that have been studied longitudinally across $T$ time points. Each time series component $Y(t)$ is assumed to be sampled $n$ times. Thus, $Y_{ij}(t)$ represents the value of the $j$-th variable at time $t$ for the $i$-th sample, $i = 1, \ldots, n$, $j = 1, \ldots, p$. 

Here, we focus on non-Gaussian multivariate time series data such as ordinal-valued time series taking values in $\{0, 1, \ldots, (c_k - 1)\}$, where $c_k$ is the number of possible categories, or mixed categorical and continuous time series data, which routinely occurs in real world settings.

\subsection{Gaussian Copula}
\label{GCGM}
To model dependencies among p-dimensional vector $y$ we use a Gaussian copula, defined as 
\begin{align}
\label{GC}
F \Big( y_1, \ldots, y_p \Big)  = & \Phi_{p} \Big( \Phi^{-1}\big( F_1(y_1)\big), \ldots , \Phi^{-1} ( F_p(y_p))  \Big  |  \Omega_{p \times p} \Big) 
\end{align}
where  $\Phi_{p}(. | \Omega)$ is the a $p$-dimensional Gaussian cdf with correlation matrix $\Omega_{p \times p}$, and $y = (y_1, \ldots, y_p)$. From equation (\ref{GC}) the following properties are clear: the joint marginal distribution of any subset of $Y$ has a Gaussian copula with a correlation matrix $\Omega$ and univariate marginals $F_j$.
The Gaussian copula can be expressed in terms of a latent Gaussian variable ${\bf Z} =  Z_1, \ldots, Z_p$ as follow
\[
\textbf{Z} \sim \mathcal{N}(0, \Omega_{p \times p} ) 
\]
and 
\begin{equation}
Y_{j} = F^{-1}_{j}( \Phi(Z_{j})).
\end{equation}

Since the marginal distributions $F_j$ are nondecreasing, observing $y_{i_1j} < y_{i_2j}$ implies $z_{i_1j} < z_{i_2j}$. This can be written as set $\mathcal{A}(y)$ where given the observed data $y_j = (y_{1,j}, \ldots, y_{n,j})$, the latent samples $z_j = (z_{1,j}, \ldots, z_{n,j})$, are constrained to belong to the set
\[
\mathcal{A}(y) = \{ z \in R^{n \times p}: \max\{z_{s,j}: y_{s,j} < y_{r,j}\} < z_{r,j} < \min\{z_{s,j}: y_{r,j} < y_{s,j} \}   \}
\] 
If an observed value of $y_j$ is missing, we define the lower bound and the upper bound of $z_j^{(r)}$ as $-\infty$ and $\infty$, respectively.  
\subsection{Model definition}
\label{likelihood}
We assume a stable dynamic chain graph model meaning that the structure of interactions within each time point remains stable for previous and current time step, and interactions between consecutive time steps are stable too. We use a vector autoregressive process of order $1$, VAR($1$), 
\begin{equation}
\label{model}
Z_t = \Gamma Z_{(t-1)} + \epsilon_t
\end{equation}
to describe the directed latent interactions, where $\epsilon_t \sim N(0, \Theta^{-1})$ describes the undirected instantaneous interactions.

The parameter set of this model contains all the conditional independence relationships in the dynamic chain graph model where the following terms hold: $\theta_{j\acute{j}} = 0$ if and only if $Z_j^{(t)} \bigCI Z_{\acute{j}}^{(t)} \ | \ Z_{- j,\acute{j}}^{(t)} Z^{(t-1)}$, and $\gamma_{j\acute{j}} = 0$ if and only if $Z_j^{(t)} \bigCI Z_{\acute{j}}^{(t-1)} \ | \ Z_{-j}^{(t)} Z_{-\acute{j}}^{(t-1)} $.

Given the set $\mathcal{A}(y)$, 
we calculate the likelihood as 
\begin{align}
\label{ERL}
f( \mathbf{y} \ | \ \Theta, \Gamma, F )   = & f(\mathbf{y}, \mathbf{z} \in \mathcal{A}(\mathbf{y}) \ | \ \Theta, \Gamma,  F ) \nonumber \\
= & f_Z( \mathbf{z} \in \mathcal{A}(\mathbf{y}) \ | \ \Theta, \Gamma) f(\mathbf{y} \ | \ \mathbf{z} \in \mathcal{A}(\mathbf{y}), \Theta, \Gamma, F)
\end{align}
where $ y = \{(y^{(t)}_1, \ldots, y^{(t)}_p)\}_{t=1}^T$ and $F = \{(F^{(t)}_1, \ldots, F^{(t)}_p)\}_{t=1}^T$. Given the set of parameters, the event $\mathbf{z} \in \mathcal{A}(\mathbf{y})$ in (\ref{ERL}) does not depends on marginals and contains the relevant information about the copula and the parameters of interest $\Theta$ and $\Gamma$. We drop the second term in (\ref{ERL}) because this term does not provide any information about intra and inter time-slice dependencies. As \cite{hoff2007extending} proposes we use $f_Z( \mathbf{z} \in \mathcal{A}(\mathbf{y}) \ | \ \Theta, \Gamma)$ as the rank likelihood, 
{\small
\begin{align}
\label{LL}  
\ell_Y(\Theta, \Gamma )= & \sum\limits_{i = 1}^{n} \log f(\mathbf{z_i} \in \mathcal{A}(\mathbf{y}) \ | \  \Theta, \Gamma ) \nonumber  \\ 
=& \sum\limits_{i = 1}^{n} \sum\limits_{t= 2}^{T} \log f(z_i^{(t)} \in \mathcal{A}(\mathbf{y}_i^{(t)}) \ | \ z_i^{(t-1)} \in \mathcal{A}(\mathbf{y}_i^{(t-1)}); \Theta, \Gamma) + \log f(z_{i}^{(1)} \in \mathcal{A}(\mathbf{y}_i^{(1)})\ | \ \Theta, \Gamma) 
\end{align}
}
We ignore the second term in (\ref{LL}) as we do not want to make additional assumption on the unconditional distribution of $Y^{(1)}$. And we start from $t = 2$, where we compute the conditional log-likelihood using the conditional distribution $f(z^{(t)} | z^{(t-1)})$. 
According to (\ref{model}) the conditional distribution $Z^{(t)} \ | \ Z^{(t-1)}$ follows a multivariate normal distribution 
\begin{equation} 
\label{condDist}
Z^{(t)} \ | \ Z^{(t-1)}= z^{(t-1)} \sim \mathcal{N}( \Gamma z^{(t-1)}, \Theta^{-1} ) 
\end{equation}
which its density for $t$-th observation is defined as 
\begin{align}
f(z^{(t)} \ | \ z^{(t-1)} ; \Theta, \Gamma ) = (2\pi)^{p/2} \det(\Theta)^{1/2} \exp\Big[ \frac{1}{2} \Big( z^{(t)} - \Gamma z^{(t-1)} \Big)' \Theta \Big( z^{(t)} - \Gamma z^{(t-1)} \Big) \Big].
\end{align}
\subsection{Penalized EM inference}
\label{Inference}
In Gaussian copula, we treat the marginals distributions as nuisance parameters since our main goal is to learn the dependence structure among time series components both at a fixed time step $t \in \mathbb N$ and also across consecutive time steps. 
We use an empirical marginal cdf $\widehat{F}_j = \frac{n}{n + 1} \sum\limits_{i = 1}^{n} \frac{1}{n} 1( y_{ij} \le y )$ \citep{genest1995semiparametric} to estimate marginals. 

Genetic time series data often are high dimensional due to a large number of variables that are measured on small number of samples across only few time steps. Furthermore, many real-world networks (e.g. genetic, genomics, and brain networks) are intrinsically sparse. Thus, incorporating sparsity into the proposed dynamic chain graph model makes the derived model more biologically plausible. Accordingly, we propose a dynamic chain graph model for genetic data based on the penalized likelihood. In order to find the penalized maximum likelihood estimation we will use the EM algorithm \citep{green1990use}. This modeling technique provides sparse estimates of the autoregressive coefficient matrix $\Gamma$ and the precision matrix $\Theta$ in (\ref{model}) which are used to reconstruct inter and intra time-slice conditional independences, respectively. 

The E-step of the EM algorithm is given by
\begin{align}
Q(\Theta, \Gamma \ | \ \Theta^\star, \Gamma^\star) = & E_z \Big[ \ell_{Y,Z}(\Theta, \Gamma) \ \Big | \ y_i, \Theta^\star, \Gamma^\star \Big] \nonumber \\
= & E_z \Big[  \sum\limits_{i = 1}^{n} \sum\limits_{t = 2}^{T} \log f(Z_i^{(t)} \ | \  Z_i^{(t-1)}; \Theta, \Gamma) \ \Big| \ y_i, \Theta^\star, \Gamma^\star \Big].
\end{align}
Under the assumption described in (\ref{condDist}), the E-step can be written as 
\begin{equation}
\label{Qfun}
Q(\Theta, \Gamma | \Theta^\star, \Gamma^\star) = \frac{n(T-1)}{2} \Big[ - p \log(2\pi) +  \log \det(\Theta) - \mbox{tr} \Big( E(S_\Gamma \ | \ y_i, \Theta^\star, \Gamma^\star) \Theta \Big) \Big]
\end{equation}
where 
\begin{align}
\label{SGamma}
E(S_\Gamma \ | \ y_i, \Theta^\star, \Gamma^\star) = & \frac{1}{n(T-1)} \sum\limits_{i=1}^{n}\sum\limits_{t=2}^{T}E_Z \Big[ (Z_i^{(t)} - \Gamma Z_i^{(t-1)}) (Z_i^{(t)} - \Gamma Z_i^{(t-1)})' \ \Big| \ y_i, \Theta^* ,\Gamma^* \Big] \nonumber \\
= & \frac{1}{n(T-1)} \Big[ S_{cc} -  S_{cp} \Gamma' - \Gamma S_{cp}' + \Gamma S_{pp} \Gamma'    \Big]
\end{align}
such that conditional expectation at current time, $S_{cc}$, and at past, $S_{pp}$, is defined as 
\[
S_{cc} =\sum\limits_{i=1}^{n}\sum\limits_{t=2}^{T}E_Z[ Z_i^{(t)} Z_i^{(t)\prime}  | y_i; \Theta^\star, \Gamma^\star] , \qquad
S_{pp} =\sum\limits_{i=1}^{n}\sum\limits_{t=1}^{T-1}E_Z[ Z_i^{(t)} Z_i^{(t)\prime}  | y_i; \Theta^\star, \Gamma^\star]
\]
and the conditional expectation at inter time-slice dependence is
\[
S_{pc} =\sum\limits_{i=1}^{n}\sum\limits_{t=2}^{T}E_Z[ Z_i^{(t-1)} Z_i^{(t)\prime}  | y_i; \Theta^\star, \Gamma^\star].
\]

The latent variables $Z_{i}^{(t-1)} = \{Z^{(t-1)}_{i,1}, \ldots, Z^{(t-1)}_{i,p}\}$ and $Z_{i}^{(t)} = \{Z^{(t)}_{i,1}, \ldots, Z^{(t)}_{i,p}\}$ is used to calculate the conditional expectation of intra time-slice dependencies $S_{pp}$ and $S_{cc}$, respectively. And $Z^{(pc)}_i = \{Z^{(t-1)}_{i,1}, \ldots, Z^{(t-1)}_{i,p}, Z^{(t)}_{i,1}, \ldots, Z^{(t)}_{i,p}\}$ is used to calculate $S_{pc}$. All the three above mentioned conditional expectations are a $p \times p$ matrix. When $j = j'$ they can be computed through the second moment $E(Z^{(t)^2}_{ij} \ | \ y_i; \Theta^\star, \Gamma^\star)$. When $j \ne j'$ we use a mean field theory approach \citep{chandler1987introduction} to approximate them as
\begin{equation}
\label{MeanField-intraTime}
E\Big( Z^{(t)}_{i,j} Z^{(t)}_{i,j'} \ \Big| \ y_i; \Theta^\star, \Gamma^\star \Big) \approx  E\Big( Z^{(t)}_{i,j}  \Big| y_i; \Theta^\star, \Gamma^\star \Big) E\Big( Z^{(t)}_{i,j'} \Big| y_i; \Theta^\star, \Gamma^\star \Big)
\end{equation}
for intra time-slice dependencies, and for inter time-slice dependencies follows as
\begin{equation}
\label{MeanField-interTime}
E\Big( Z^{(t-1)}_{i,j} Z^{(t)}_{i,j} \Big| y_i; \Theta^\star, \Gamma^\star \Big) \approx  E\Big( Z^{(t-1)}_{i,j}  \Big|  y_i; \Theta^\star, \Gamma^\star \Big) E\Big( Z^{(t)}_{i,j'} \Big| y_i; \Theta^\star, \Gamma^\star \Big)
\end{equation}
This approximation performs well when the interaction between $Z^{(t)}_{i,j}$ and $Z^{(t)}_{i,j'}$ given the rest of the variables, and the interaction between $Z^{(t -1)}_{i,j}$ and $Z^{(t)}_{i,j'}$ given the rest of the variables are close to be independent; this often holds in our proposed dynamic chain graph model which $\Theta$ and $\Gamma$ are sparse. 

When $j \ne j'$ the off-diagonal elements of $S_{cc}$, $S_{pp}$, and $S_{pc}$ matrices can be computed through the first moment as
\begin{equation}
\label{exactFirstMoment}
E \Big( Z^{(t)}_{i,j}  \ \Big| \ y_i; \Theta^\star, \Gamma^\star \Big) = E \Big[ E \Big(Z^{(t)}_{i,j} | Z^{(t-1)}_i, Z^{(t)}_{i,-j}, Z^{(t+1)}_i, y^{(t)}_{i,j}; \Theta,\Gamma \Big) \Big| y_i; \Theta^\star,\Gamma^\star \Big]  
\end{equation}
and the second moments is 
\begin{equation}
\label{exactSecondMoment}
E\Big(Z^{(t)^2}_{i,j} | y_i; \Theta^\star, \Gamma^\star \Big) = E \Big[ E \Big(Z^{(t)^2}_{i,j} | Z^{(t-1)}_i, Z^{(t)}_{i,-j}, Z^{(t+1)}_{i}, y^{(t)}_{i,j}; \Theta, \Gamma \Big) \Big| y_i; \Theta^\star \Gamma^\star \Big]  
\end{equation}
Given the property of Gaussian distribution, $(Z^{(t)}_i, Z_i^{(t+1)}) \ | \ Z^{(t-1)}_i ; \Theta, \Gamma$ follows a multivariate normal distribution with mean and variance-covariance matrix  
\[\mu = \begin{bmatrix} 
\Gamma z^{(t-1)}_i  \\
\Gamma^2 z^{(t-1)}_i  
\end{bmatrix} 
\qquad \qquad \qquad
V = \begin{bmatrix} 
\Theta^{-1} & \Theta^{-1} \Gamma \\
\Gamma \Theta^{-1} & \Gamma \Theta^{-1} \Gamma'+ \Theta^{-1} 
\end{bmatrix}.
\]
Therefore, the conditional distribution of $Z^{(t)}_{i,j} \ | \ Z^{(t-1)}_i, Z^{(t)}_{i,-j}, Z^{(t+1)}_i ; \Theta, \Gamma$ inside the inner expectation in (\ref{exactFirstMoment}) and (\ref{exactSecondMoment}) follows a multivariate normal distribution with mean $\mu_{ij}$ and variance $v_{ij}$ as follow
\[
\mu_{ij} = ( \Gamma_i z^{(t-1)}_i )_j + V_{j,-j} V^{-1}_{-j,-j} \Big(  \begin{bmatrix} 
 z^{(t)}_{i,-j}  \\
 z^{(t+1)}_{i}  
\end{bmatrix} - \begin{bmatrix} 
\Gamma z^{(t-1)}_i  \\
\Gamma^2 z^{(t-1)}_i  
\end{bmatrix}  \Big)
\]
\[
v_{ij} = V_{j,j} - V_{j,-j} V^{-1}_{-j,-j} V_{-j,j}.
\]
Calculating the exact value of the first and second moments is computationally expensive. Moreover, we 
approximate the first and the second moments as follow
\begin{equation}
\label{firstMoment:inter}
E \Big( Z^{(t)}_{i,j}  \Big| y_i; \Theta^\star, \Gamma^\star \Big) \approx E \Big[ E \Big(Z^{(t)}_{i,j} | Z^{(t-1)}_i, Z^{(t)}_{i,-j}, y^{(t)}_{i,j}; \Theta, \Gamma \Big) \Big| y_i; \Theta^\star, \Gamma^\star, \Big]  
\end{equation}
\begin{equation}
\label{secondMoment:inter}
E\Big(Z^{(t)^2}_{i,j} \ | \ y_i; \Theta^\star, \Gamma^\star \Big) \approx E \Big[ E \Big(Z^{(t)^2}_{i,j} \ | \ Z^{(t-1)}_i, Z^{(t)}_{i,-j}, y^{(t)}_{i,j}; \Theta, \Gamma \Big) \ \Big| \ y_i; \Theta^\star , \Gamma^\star\Big]  
\end{equation}

The conditional distribution of $Z^{(t)}_{i} | Z^{(t-1)}_i ; \Theta, \Gamma$ follows a multivariate normal distribution with mean $\Gamma_i z^{(t-1)}_i$ and variance-covariance matrix $\Theta^{-1}$. Due to a property of Gaussian distribution, the conditional distribution of $Z^{(t)}_{i,j} | Z^{(t-1)}_i, Z^{(t)}_{i,-j} ; \Theta, \Gamma;$ inside the inner expectation in (\ref{firstMoment:inter}) and (\ref{secondMoment:inter}) follows a multivariate normal distribution with mean and variance-covariance matrix as follow
\[
\mu'_{i,j} = (\Gamma_i z^{(t-1)}_i)_j + \widehat{\Sigma}_{j,-j} \widehat{\Sigma}^{-1}_{-j,-j} \Big( z^{(t)\tau}_{i,-j} -(\Gamma_i z_i^{(t-1)})_{-j} \Big)
\]
\[
\sigma'^2_{i,j} = \widehat{\Sigma}_{j,j} - \widehat{\Sigma}_{j,-j} \widehat{\Sigma}^{-1}_{-j,-j}\widehat{\Sigma}_{-j,j}.
\]

We remark that conditioning $z^{(t)}_{i,j}$ on $z^{(t-1)}_{i}$, $z^{(t)}_{i,-j}$ and $y^{(t)}_{i,j}$ is equivalent to 
\[z^{(t)}_{i,j} | z^{(t-1)}_{i}, z^{(t)}_{i,-j}, c_{j, y^{(t)}_{ij}} \le z^{(t)}_{ij} \le c_{j, y^{(t)}_{ij}+1}.
\]
Thus, this conditional distributions follows a truncated normal on the interval $[c_{j, y^{(t)}_{ij}}, c_{j, y^{(t)}_{ij}+1}]$ which the first and second moments can be obtained via lemma \ref{lemm}.

\begin{lemma}\citep{johnson1995noncentral}.
\label{lemm}
Let $Z \sim \mathcal{N}(\mu_0, \sigma^2_0)$ such that $\delta_1 = (c_1 - \mu_0)/\sigma_0$ and $\delta_2 = (c_2 - \mu_0)/\sigma_0$ are true for any constants that $c_1 < c_2$. Then the first and second moments of the truncated normal distribution on the interval $(c_1, c_2)$ are defined as
\begin{equation}
E(Z | c_1 \leq Z \leq c_2) = \mu_0 + \frac{\phi(\delta_1) - \phi(\delta_2)}{\Phi(\delta_2) - \Phi(\delta_1)} \sigma_0 \nonumber
\end{equation}
\begin{equation}
E(Z^2 | c_1 \leq Z \leq c_2)= \mu_0^2 + \sigma_0^2 + 2 \frac{\phi(\delta_1) - \phi(\delta_2)}{\Phi(\delta_2) - \Phi(\delta_1)} \mu_0 \sigma_0 + \frac{\delta_1 \phi(\delta_1) - \delta_2 \phi(\delta_2)}{\Phi(\delta_2) - \Phi(\delta_1)} \sigma_0^2 \nonumber
\end{equation}
where $\phi(.)$ is the density function of the standard normal distribution. 
\end{lemma}

Both means $\mu_{i,j}$ and $\mu'_{i,j}$ are a linear function of $z^{(t)}_{i,-j}$, and both $\frac{\phi(\delta_1) - \phi(\delta_2)}{\Phi(\delta_2) - \Phi(\delta_1)}$ and $\frac{\delta_1 \phi(\delta_1) - \delta_2 \phi(\delta_2)}{\Phi(\delta_2) - \Phi(\delta_1)}$ are nonlinear functions of $z^{(t)}_{i,-j}$. Applying Lemma \ref{lemm} on the conditional expectations in (\ref{firstMoment:inter}) and (\ref{secondMoment:inter}) leads to following approximations 
{\footnotesize
\begin{align}
\label{offDiag:inter}
E(Z^{(t)}_{i,j} \ | \ y_i; \Theta^\star, \Gamma^\star)  \approx &
 \Sigma_{j,-j} \Sigma^{-1}_{-j,-j} \Big( E(Z^{(t)\tau}_{i,-j}\ | \ y_i; \Theta^\star, \Gamma^\star ) -(\Gamma_i E(Z^{(t-1)}_{i}\ | \ y_i; \Theta^\star, \Gamma^\star))_{-j} \Big) \nonumber \\
 & + (\Gamma_i E(Z^{(t-1)}_{i}\ | \ y_i; \Theta^\star, \Gamma^\star))_j + \frac{\phi(\delta_{i,j,y_{i,j}^{(t)}}^{(t)}- \phi(\delta_{i,j,y_{i,j}^{(t)}+1}^{(t)})}{\Phi(\delta_{i,j,y_{i,j}^{(t)}+1}^{(t)}) - \Phi(\delta_{i,j,y_{{i,j}}^{(t)}}^{(t)})} \sigma_{i,j}
\end{align}
\begin{align}
\label{diag:inter}
E((Z^{(t)^2}_{i,j}) \ | \ y_i; \Theta^\star, \Gamma^\star) \approx &  \Big( (\Gamma_i E(Z^{(t-1)}_{i}\ | \ y_i; \Theta^\star, \Gamma^\star))_j \Big)^2 + \Sigma_{j,-j} \Sigma^{-1}_{-j,-j} E(Z^{(t)\tau}_{i,-j} Z^{(t)}_{i,-j} \ | \ y_i; \Theta^\star, \Gamma^\star) \Sigma^{-1}_{-j,-j}\nonumber \\
	&\Sigma_{-j,j} + \Sigma_{j,-j} \Sigma^{-1}_{-j,-j} \Big( (\Gamma_i E(Z^{(t-1)}_{i}\ | \ y_i; \Theta^\star, \Gamma^\star))_{-j} \Big)^2 \Sigma^{-1}_{-j,-j} \Sigma_{-j,j} -2 \Sigma_{j,-j}  \nonumber \\
	&\Sigma^{-1}_{-j,-j} \quad E(Z^{(t)\tau}_{i,-j} \ | \ y_i; \Theta^\star, \Gamma^\star) \Big( (\Gamma_i E(Z^{(t-1)}_{i}\ | \ y_i; \Theta^\star, \Gamma^\star )) \Big)_{-j} \Sigma^{-1}_{-j,-j} \Sigma_{-j,j}  \nonumber \\
	&  + 2 \Big(\Gamma_i E(Z^{(t-1)}_{i}\ | \ y_i; \Theta^\star, \Gamma^\star )\Big)_j \Sigma_{j,-j} \Sigma^{(-1)}_{-j,-j}  E(Z^{(t)}_{i,-j}\ | \ y_i; \Theta^\star, \Gamma^\star)  \nonumber \\
	&  - 2 (\Gamma_i E(Z^{(t-1)}_{i}\ | \ y_i; \Theta^\star, \Gamma^\star))_j \Sigma_{j,-j} \Sigma^{(-1)}_{-j,-j} \Big(\Gamma_i E(Z^{(t-1)}_{i}\ | \ y_i; \Theta^\star, \Gamma^\star)\Big)_{-j} + \sigma^2_{i,j}  \nonumber \\
	& + 2 \frac{\phi(\delta_{i,j,y_{i,j}^{(t)}}^{(t)})- \phi(\delta_{i,j,y_{i,j}^{(t)}+1}^{(t)})}{\Phi(\delta_{i,j,y_{i,j}^{(t)}+1}^{(t)}) - \Phi(\delta_{i,j,y_{{i,j}}^{(t)}}^{(t)})} \Big[(\Gamma_i E(Z^{(t-1)}_{i}\ | \ y_i; \Theta^\star, \Gamma^\star))_j \nonumber \\
	& + \Sigma_{j,-j} \Sigma^{-1}_{-j,-j} \Big( E(Z^{(t)\tau}_{i,-j}\ | \ y_i; \Theta^\star, \Gamma^\star) -(\Gamma_i E(Z^{(t-1)}_{i}\ | \ y_i; \Theta^\star, \Gamma^\star))_{-j} \Big) \Big] \sigma_{i,j} \nonumber \\
	& + \frac{\delta^{(t)}_{i,j,y_{i,j}^{(t)}} \phi(\delta_{i,j,y_{i,j}^{(t)}}^{(t)}) - \delta_{i,j,y_{i,j}^{(t)}+1}^{(t)} \phi(\delta_{i,j,y_{i,j}^{(t)}+1}^{(t)})}{\Phi(\delta_{i,j,y_{i,j}^{(t)}+1}^{(t)}) - \Phi(\delta_{i,j,y_{i,j}^{(t)}}^{(t)})}\sigma^2_{i,j}
	\end{align} 
}
where $\delta^{(t)}_{i,j,y^{(t)}_{i,j}} = (c^{(t)}_{i,j} - \mu'_{i,j}) / \sigma' _{ij}$. Here, the first order delta method is used to approximate the nonlinear terms [more details in \citep{guo2015graphical}]. Moreover, we approximate the elements of inter time-slice conditional expectation matrix $S_{pc}$ through equations (\ref{offDiag:inter}) and (\ref{diag:inter}). For approximating the elements of intra time-slice conditional expectation matrices $S_{pp}$, $S_{cc}$ we refer to the Appendix.

The M-step of the EM algorithm contains two-stage optimization process where we maximize expectation of the penalized log-likelihood with respect to $\Theta$ and $\Gamma$. We introduce two different penalty functions $P_\lambda(.)$ and $P_\rho(.)$ for intra time-slice conditional independencies $\Theta$, and inter time-slice conditional independencies $\Gamma$, respectively. Therefore, the objective function for optimization can be defined as 
\begin{equation}
\label{objFun}
Q_{pen} (\Theta, \Gamma| \Theta^\star_\lambda, \Gamma^\star_\rho) = \frac{n(T-1)}{2} \Big[ \log \det(\Theta) - \mbox{tr} \Big( \Theta  S^{(E)}_{\Gamma} \Big) \Big] - \sum\limits_{j \ne j'}^{p} P_\lambda(|\theta_{jj'}|) - \sum\limits_{j,j'}^{p} P_\rho(|\gamma_{jj'}|)
\end{equation}
where $S^{(E)}_\Gamma$ denotes the expectation of $S_\Gamma$ given the data and updated parameters, and $\theta_{jj'}$ and $\gamma_{jj'}$ are the $jj'$-th element of the $\Theta$ and $\Gamma$ matrices. Among different penalty functions, we consider the $L_1$ norm and smoothly clipped absolute deviation (SCAD) penalty functions which have the desirable sparsity properties.

\paragraph{$L_1$ penalized EM.} The Lasso or $L_1$ penalty function is defined as 
\[
P_\lambda (\theta) = \lambda |\theta|.
\]
The $L_1$ penalty leads to a desirable optimization problem, where the log-likelihood is convex and can efficiently be solved using various optimization algorithms at the $k$-th iteration of the EM. Under this penalty function, the updated estimates are given via
\begin{equation}
\label{L1normLL}
(\Theta_\lambda^{(k)}, \Gamma_\rho^{(k)}) = \arg \max_{\Theta, \Gamma}\Big\{ \log \det(\Theta) - \mbox{tr} \Big( S^{(E)}_{\Gamma} \Theta \Big)  - \lambda \sum\limits_{j \ne j'}^{p} |\theta_{jj'}| - \rho \sum\limits_{j,j'}^{p} |\gamma_{jj'}| \Big\}
\end{equation}
where the sparsity level of intra and inter time-slice conditional independences are controlled by $\lambda$ and $\rho$. $L_1$ penalty is biased due to its constant rate of penalty. To address this issue, \cite{fan2001variable} proposed SCAD penalty, which results in unbiased estimates for large coefficients. 

\paragraph{SCAD penalized EM.} The SCAD penalty function is expressed as
\[ P_{\lambda,a}(\theta)= \left\{
\begin{array}{ll}
	\lambda | \theta | & \mbox{if $\theta \le \lambda$, } \\ \\
	- \frac{|\theta^2| - 2a \lambda |\theta| + \lambda^2}{2(a - 1)} &  \mbox{if $\lambda \le |\theta| \le a \lambda$},\\ \\
	\frac{(a + 1)^2 \lambda^2}{2} &  \mbox{if $ |\theta| > a \lambda$}.
\end{array} \right. \]  
where $\lambda$ and $a$ are two tunning parameters. The function $P_{\lambda,a}(\theta)$  corresponds to a quadratic spline on $[0, \infty)$ with knots at $\lambda$ and $a\lambda$. A similar function can be written for $P_{\rho, a} (\gamma)$ where $\rho$ and $a\rho$ are two knots. The SCAD penalty is symmetric but non-convex, whose first order derivative is given by 
\[
P'_{\lambda, a}(\theta) = \lambda \Big \{ I(|\theta| \le \lambda ) + \frac{(a\lambda - |\theta|)_+}{(a-1)\lambda} I(|\theta| > \lambda) \Big \},  \qquad a > 2
\]
The notation $z_+$ stands for the positive part of $z$. \cite{fan2001variable} showed that in practice $a = 3.7$ is a good choice. Maximizing non-convex penalized likelihood is challenging. To address this issue, we use an efficient algorithm proposed in \cite{fan2009network}, which is based on local linear approximation, to maximize the penalized log-likelihood for the SCAD penalty function. In each its step, a symmetric linear function is used to locally approximates the SCAD penalty. Using the Taylor expansion, $P_{\lambda, a}(\theta)$ and $P_{\rho, a}(\gamma)$ can be approximated in the neighbor of $\theta_0$ and $\gamma_0$ as follow: 
\[
P_\lambda(|\theta|) \approx P_\lambda(|\theta_0|) + P'_\lambda(|\lambda_0|)(|\theta| - |\theta_0|)
\]
\[
P_\rho(|\gamma|) \approx P_\rho(|\gamma_0|) + P'_\rho(|\rho|)(|\gamma| - |\gamma_0|).
\]
Due to the monotonicity of $P_\lambda(.)$ and $P_\rho(.)$ over $[0,\infty)$, the derivatives $P'_\lambda(.)= \frac{\partial}{\partial\theta}(P_\lambda(\theta))$ and $P'_\rho(.)=\frac{\partial}{\partial\gamma}(P_\rho(\gamma))$ are non-negative for $\theta \in [0, \infty)$ and $\gamma \in [0, \infty)$. Therefore, under the penalized log-likelihood with SCAD penalty, the estimate of the sparse parameters $\Theta^{(k)}$ and $\Gamma^{(k)}$ relies on the solution of the following optimization problem at step $k$

\begin{equation}
\label{SCADLL}
(\Theta^{(k)}_\lambda, \Gamma_\rho^{(k)}) = \arg\max_{\Theta, \Gamma} \Big\{ \log\det(\Theta) - tr \Big( S^{(E)}_{\Gamma} \Theta \Big) - \sum\limits_{j \ne j'}^{p} w_{jj'}|\theta_{jj'}| - \sum\limits_{j,l}^{p} \nu_{jl}|\gamma_{jl}| \Big\}
\end{equation}

where $w_{jj'} = P'_\lambda(\theta^{(k)}_{jj'})$, $\nu_{jl} = P'_\rho(\gamma^{(k)}_{jl})$, and $\theta^{(k)}_{jj'}$,  $\gamma^{(k)}_{jl}$ are $jj'$-th element of $\Theta$ and $jl$-th element of $\Gamma$, respectively. The SCAD penalty applies a constant penalty to large coefficients, whereas the $L_1$ penalty increases linearly as $|\theta|$ increases. This features keep the SCAD penalty against producing biases for estimating large coefficients. Therefore, the SCAD penalty overcome the bias issue of the $L_1$ penalty. Then a two stage-optimization problem within the M-step of the EM algorithm is employed to solve the objective functions (\ref{L1normLL}) or (\ref{SCADLL}) to estimate the parameters $\Theta$ and $\Gamma$.  

\paragraph{Glasso calculation of $\Theta^{(k)}$.} For the SCAD penalty-based estimation, in the first stage we optimize
\[
\Theta^{(k)}_\lambda = \arg \max _ \Theta \Big\{ \log \det (\Theta) - \mbox{tr} (S^{(E)}_{\Gamma^\star} \Theta) - \sum\limits_{j \ne j'}^{p} w_{jj'}|\theta_{jj'}| \Big\},
\]
for previous $\Gamma^\star$. This optimization can be solved efficiently using the graphical lasso algorithm proposed by \cite{friedman2008sparse}. Due to the sparsity in each iteration, we consider a one-step local linear approximation algorithm (LLA). \cite{zou2008one} showed that one-step LLA, asymptotically, performs as well as the fully iterative LLA algorithm as long as initial solution is good enough. In practice, we take the initial value as the $L_1$ penalty graphical LASSO for estimating the intra time-slice conditional independences $\Theta$ in order to calculate the initial weights $w_{jj'}$ and $\nu_{jl}$.

\paragraph{Regularized coordinate descent algorithm for $\Gamma^{(k)}$.} After we finish an updating $\Theta$ in the first-stage of the optimization, in the second-stage we proceed to update the estimate of $\Gamma$ given the updated $\Theta$. In the SCAD penalty-based we optimize 

{\small
\begin{align}
\label{GammaSCAD}
\Gamma_\rho^{(k)} & = \arg  \max _ \Gamma \Big\{ \log \det (\Theta_\lambda^{(k)}) - \mbox{tr} (S^{(E)}_{\Gamma} \Theta_\lambda^{(k)}) - \sum\limits_{j,l}^{p} \nu_{jl}|\gamma_{jl}| \Big\}  \nonumber \\
& =  \arg \max_\Gamma \Big\{\log \det (\Theta_\lambda^{(k)}) - \mbox{tr} (S_{cc} \Theta_\lambda^{(k)} -  S_{cp} \Gamma' \Theta_\lambda^{(k)} - \Gamma S_{cp}'\Theta_\lambda^{(k)} + \Gamma S_{pp} \Gamma' \Theta_\lambda^{(k)}) - \sum\limits_{j,l}^{p} \nu_{jl}|\gamma_{jl}| \Big\}.
\end{align} 
}

This objective function is quadratic in $\Gamma$ for given $\Theta_\lambda^{(k)}$. Thus, we use a direct coordinate descent algorithm to calculate $\Gamma_\rho^{(k)}$. So, the derivative of the penalized negative log-likelihood (\ref{GammaSCAD}) with respect to $\gamma_{jl}$ is 
\begin{equation}
\label{derivative.wrt.Gamma}
\frac{\partial \ell_p}{\partial \gamma_{jl}} = -2 e'_j(S'_{cp} \Theta_\lambda^{(k)}) e_i + 2 e'_j(S_{cc} \Gamma' \Theta_\lambda^{(k)}) e_i + \nu_{jl} \mbox{sgn}(\gamma_{jl})
\end{equation}
where $\mbox{sgn}(.)$ is the sign function. These are the Karush–Kuhn–Tucker (KKT) equations defining the solution to the maximization problem. We note that for an arbitrary matrix $A_{p \times p}$, $\partial\mbox{tr}(\Gamma A)/\partial\gamma_{jl}= a_{lj} = e'_l A e_j$, where $e_l$ and $e_j$ are the corresponding base vector with $p$ dimension each. Setting the derivative of negative log-likelihood (\ref{derivative.wrt.Gamma}) to zero, we get an update for the elements of $\Gamma$ matrix as follow
\begin{equation}
\label{gammaHat}
\gamma_{jl} = \mbox{sgn}(g_{jl}) \frac{(|g_{jl}| - \nu_{jl})_+}{2(e'_l S_{cc} e_l )(e'_j \Theta_\lambda^{(k)} e_j)},
\end{equation}
where $g_{jl} = 2\{ e'_l(S'_{cp} \Theta_\lambda^{(k)}) e_j +(e'_l S_{cc} e_l)(e'_j \Theta_\lambda^{(k)} e_j) \gamma_{jl} - e'_l(S_{cc} \Gamma'\Theta_\lambda^{(k)})e_j\}$, $\gamma_{jl}$, and $\Gamma_\rho^{(k)}$ are the estimates in the last step of the iteration inside the optimization (\ref{gammaHat}).

Given the two-stage optimization problem inside the M-step, we update the $S_\Gamma$ matrix in the E-step. This iterative procedure continues until the difference between previous $(\Theta_\lambda^{(k - 1)}, \Gamma_\rho^{(k - 1)})$ and updated $(\Theta_\lambda^{(k)}, \Gamma_\rho^{(k)})$ becomes smaller than a, user specified, tolerance. Based on our simulation experiments, the EM algorithm converges in a few iterations (at most $5$ iterations is needed to reach the convergence). We define the estimate as the stationary point of the EM, $(\widehat{\Theta}_\lambda, \widehat{\Gamma}_\rho) = \lim\limits_{k \rightarrow \infty} (\Theta_\lambda^{(k)}, \Gamma_\rho^{(k)})$.

\renewcommand{\tabcolsep}{5pt} 
\begin{table}[H]  \footnotesize 
	\vspace{0em} 
	\caption{Performance measure results of the simulation study for tsnetwork and SparseTSCGM using SCAD penalized likelihood estimation for the precision and autoregressive coefficient matrices for fixed time point, t=5. In SparseTSCGM* the normal transformation is applied to the simulated ordinal data. \label{tablet5} }
	\centering
	\begin{tabular}{l*{8}{l}l}
		\toprule
		\!&       \!& \multicolumn{3}{c}{Performance $\Theta$}                                & \multicolumn{4}{c}{Performance $\Gamma$}                            \\
		\cmidrule{3-5}                                                \cmidrule{7-9}      
		Fixed at t= 5 &     \!& \!$F_1$ score    & SEN    & SPE           \!&\!&\!$F_1$ score  \!&\!SEN        \!&\!SPE         \\
		\midrule        
		\label{sim}
		\\
		\multicolumn{2}{l}{p=10 \& n=20 }                                           \\
		tsnetwork 		 &   \!&\!0.35 & 0.35    & 0.77    \!&\!&\!0.42    \!&\!0.43    \!&\!0.68  \\
		SparseTSCGM  &   \!&\!0.14 & 0.14    & 0.89    \!&\!&\!0.42    \!&\!0.67    \!&\!0.34 \\
		SparseTSCGM* &   \!&\!0.20    & 0.18    & 0.88    \!&\!&\!0.40       \!&\!0.47       \!&\!0.56     \\
		\\
		\multicolumn{2}{l}{p=10 \& n=50}    \\                           
		tsnetwork 			&		\!&\!0.37 & 0.37  & 0.85   \!&\!&\!0.44    \!&\!0.43    \!&\!0.7  \\
		SparseTSCGM 	&		\!&\!0.33 & 0.45  & 0.80   \!&\!&\!0.42    \!&\!0.65    \!&\!0.34 \\
		SparseTSCGM*	&       \!&\!0.31	  & 0.32  & 0.86   \!&\!&\!0.42       \!&\!0.45       \!&\!0.63    \\
		
		\\
		\multicolumn{2}{l}{p=50 \& n=20 }    \\                           
		tsnetwork 		&   \!&\!0.18 & 0.12     & 0.98    \!&\!&\!0.30 \!&\!0.30 \!&\!0.93  \\
		SparseTSCGM &   \!&\!0.02 & 0.03     & 0.95    \!&\!&\!0.31 \!&\!0.54 \!&\!0.81  \\
		SparseTSCGM*&   \!&\!0 .00   & 0.00  & 1.00    \!&\!&\!0.31 \!&\!0.22 \!&\!0.98  \\
		
		\\
		\multicolumn{2}{l}{p=50 \& n=50 }    \\                           
		tsnetwork 		&   \!&\!0.13    & 0.08  & 1.00    \!&\!&\!0.32    \!&\!0.24    \!&\!0.95  \\
		SparseTSCGM &   \!&\!0.03    & 0.03  & 0.97    \!&\!&\!0.33    \!&\!0.55    \!&\!0.82  \\
		SparseTSCGM*&   \!&\!0.07    & 0.04  & 1.00    \!&\!&\!0.28    \!&\!0.25    \!&\!0.92  \\	
		\\				
		\hline  
	\end{tabular}
\end{table}

\subsection{Selection of tuning parameters}
\label{modelSelection}
To determine the sparsity of the proposed dynamic chain graph model, the tunning parameters $\lambda$ and $\rho$ have to be tuned. We focus on estimating the sparse intra and inter time-slice conditional independences $\Theta$ and $\Gamma$, we employ the Bayesian information criteria (BIC) 

\begin{align}
\footnotesize
\label{BIC}
\mbox{BIC}(\lambda, \rho) & = -2 \ell_Y(\widehat{\Theta}_\lambda, \widehat{\Gamma}_\rho) + \log(n(T-1)) \Big(\mbox{df}(\widehat{\Theta}_\lambda) /2 + df(\widehat{\Gamma}_\rho) + p \Big) \nonumber \\ 
& \approx  n (T-1) \Big\{ \log(\det(\widehat{\Theta}_\lambda) - \mbox{tr}(S^{(E)}_{\widehat{\Gamma}_\rho} \widehat{\Theta}_\lambda) \Big\} + \log(n(T-1)) \Big(\mbox{df}(\widehat{\Theta}_\lambda) /2 + df(\widehat{\Gamma}_\rho) + p \Big)
\end{align}

to select the tuning parameters $\lambda$ and $\rho$, where $T$ and $p$ are the number of time points and the number of variables, respectively, and $\mbox{df}(\widehat{\Theta}_\lambda)$ shows the number of non-zero elements in the off-diagonal of $\widehat{\Theta}_\lambda$, and $\mbox{df}(\widehat{\Gamma}_\rho)$ is the number of non-zero elements of $\widehat{\Gamma}_\rho$. The approximation made in BIC is the result of a Laplace-type of approximation, which makes fast calculation feasible. We choose the optimal value of the penalty parameters that minimizes $BIC(\lambda, \rho)$ on a grid of candidate values for $\lambda$ and $\rho$. One may consider other information criteria that suits for graph estimations. \cite{wang2007tuning} and \cite{yin2011sparse} has been shown that BIC performs well for selecting the tunning parameter of penalized likelihood estimation.


\section{Simulation study}
\label{Sim}
To investigate and assess the performance of the proposed dynamic chain graph model, we set up a simulation to generate sparse $\Theta$ and $\Gamma$ matrices similar to \cite{abegaz2013sparse}, and \cite{yin2011sparse}. Here we evaluate the performance of the proposed method with respect to different random graph structures for $\Theta$ and $\Gamma$ matrices. Simulating different graph structures for $\Theta$ can be performed through the R package \emph{flare}. For generating $\Gamma$ matrix we took the upper diagonal of an independently generated $\Theta$ a long with a $0.2\%$ nonzero diagonal elements sampled from uniform $(0,1)$, similar to the R package \emph{SparseTSCGM}.

\renewcommand{\tabcolsep}{4pt} 
\begin{table}[H]  \footnotesize 
	\vspace{0em} 
	\caption{ Performance measure results of the simulation study for tsnetwork and SparseTSCGM using SCAD penalized likelihood estimation for the precision and autoregressive coefficient matrices for fixed time point, t=10. In SparseTSCGM* the normal transformation is applied to the simulated ordinal data. \label{tablet10} }
	\centering
	\begin{tabular}{l*{8}{l}l}
		\toprule
		\!&       \!& \multicolumn{3}{c}{Performance $\Theta$}                                & \multicolumn{4}{c}{Performance $\Gamma$}                            \\
		\cmidrule{3-5}                                                \cmidrule{7-9}      
		Fixed at t= 10 &     \!& \!$F_1$ score    & SEN    & SPE           \!&\!&\!$F_1$ score  \!&\!SEN        \!&\!SPE         \\
		\midrule        
		\label{sim}
		\\
		\multicolumn{2}{l}{p=10 \& n=20 }                                           \\
		tsnetwork 		 &   \!&\!0.35 & 0.35    & 0.77    \!&\!&\!0.43    \!&\!0.43    \!&\!0.68  \\
		SparseTSCGM  &   \!&\!0.23 & 0.32    & 0.76    \!&\!&\!0.40    \!&\!0.61    \!&\!0.34  \\
		SparseTSCGM* &   \!&\!0.26    & 0.27    & 0.88    \!&\!&\!0.41       \!&\!0.46       \!&\!0.57     \\
		\\
		\multicolumn{2}{l}{p=10 \& n=50}    \\                           
		tsnetwork 			&		\!&\!0.38 & 0.37  & 0.85   \!&\!&\!0.44    \!&\!0.43    \!&\!0.7  \\
		SparseTSCGM 	&		\!&\!0.40 & 0.59  & 0.69   \!&\!&\!0.41    \!&\!0.64    \!&\!0.32 \\
		SparseTSCGM*	&       \!&\!0.36	  & 0.40  & 0.86   \!&\!&\!0.43       \!&\!0.47       \!&\!0.61    \\
		
		\\
		\multicolumn{2}{l}{p=50 \& n=20 }    \\                           
		tsnetwork 		&   \!&\!0.11 & 0.07     & 0.99    \!&\!&\!0.31 \!&\!0.26 \!&\!0.95  \\
		SparseTSCGM &   \!&\!0.02 & 0.02     & 0.98    \!&\!&\!0.33 \!&\!0.55 \!&\!0.77  \\
		SparseTSCGM*&   \!&\!0.05 & 0.03     & 1.00    \!&\!&\!0.29 \!&\!0.25 \!&\!0.93   \\
		
		\\
		\multicolumn{2}{l}{p=50 \& n=50 }    \\                           
		tsnetwork 		&   \!&\!0.37    & 0.30  & 0.98    \!&\!&\!0.31    \!&\!0.25    \!&\!0.95  \\
		SparseTSCGM &   \!&\!0.39    & 0.34  & 0.99    \!&\!&\!0.24    \!&\!0.67    \!&\!0.64  \\
		SparseTSCGM*&   \!&\!0.34    & 0.35  & 0.97    \!&\!&\!0.28    \!&\!0.26    \!&\!0.92     \\	
		\\				
		\hline 
	\end{tabular}
\end{table}

\begin{figure}
	\centering
	\includegraphics[width=0.48\textwidth]{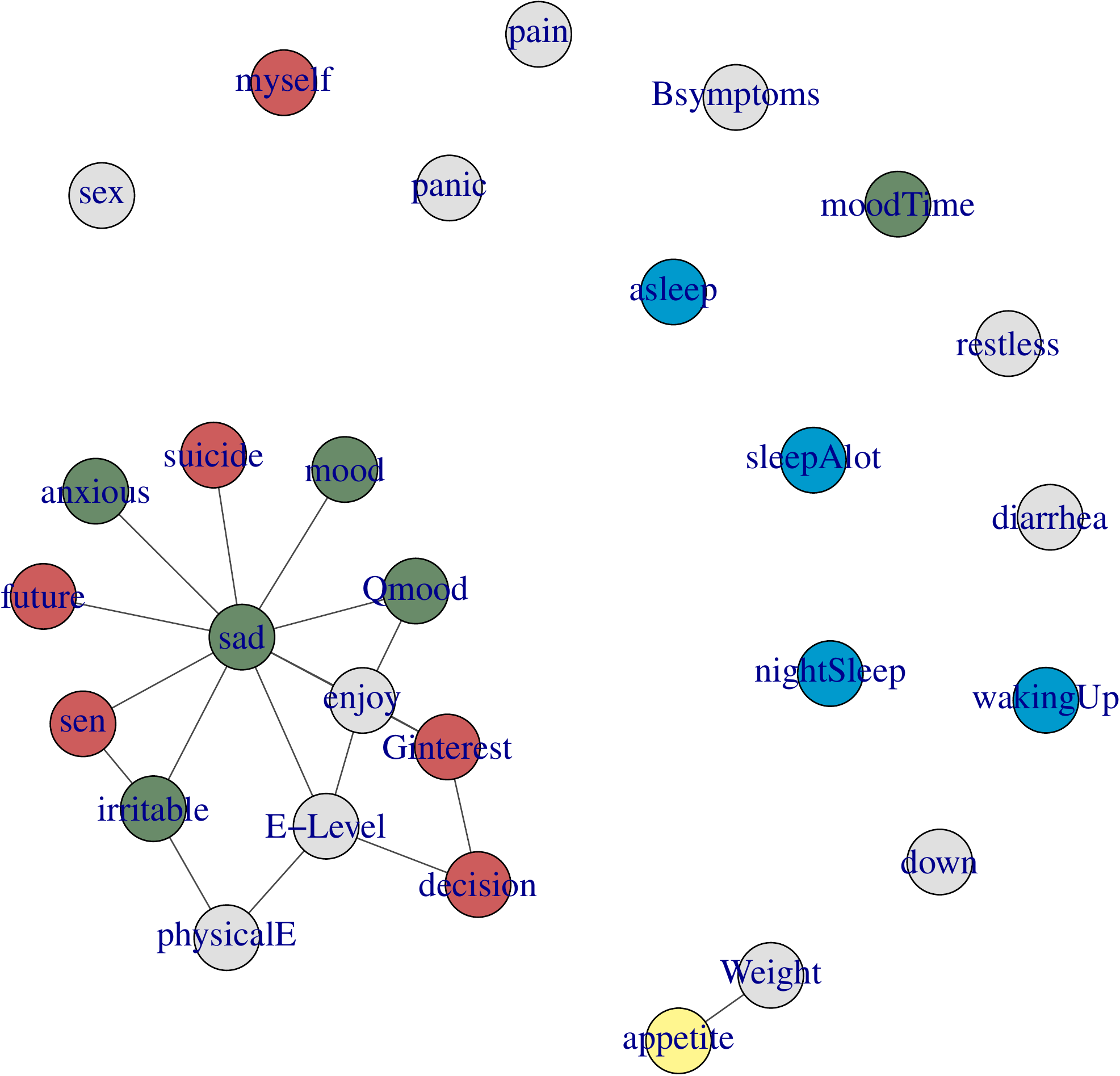} \hspace{0.1cm}
	\includegraphics[width=0.49\textwidth]{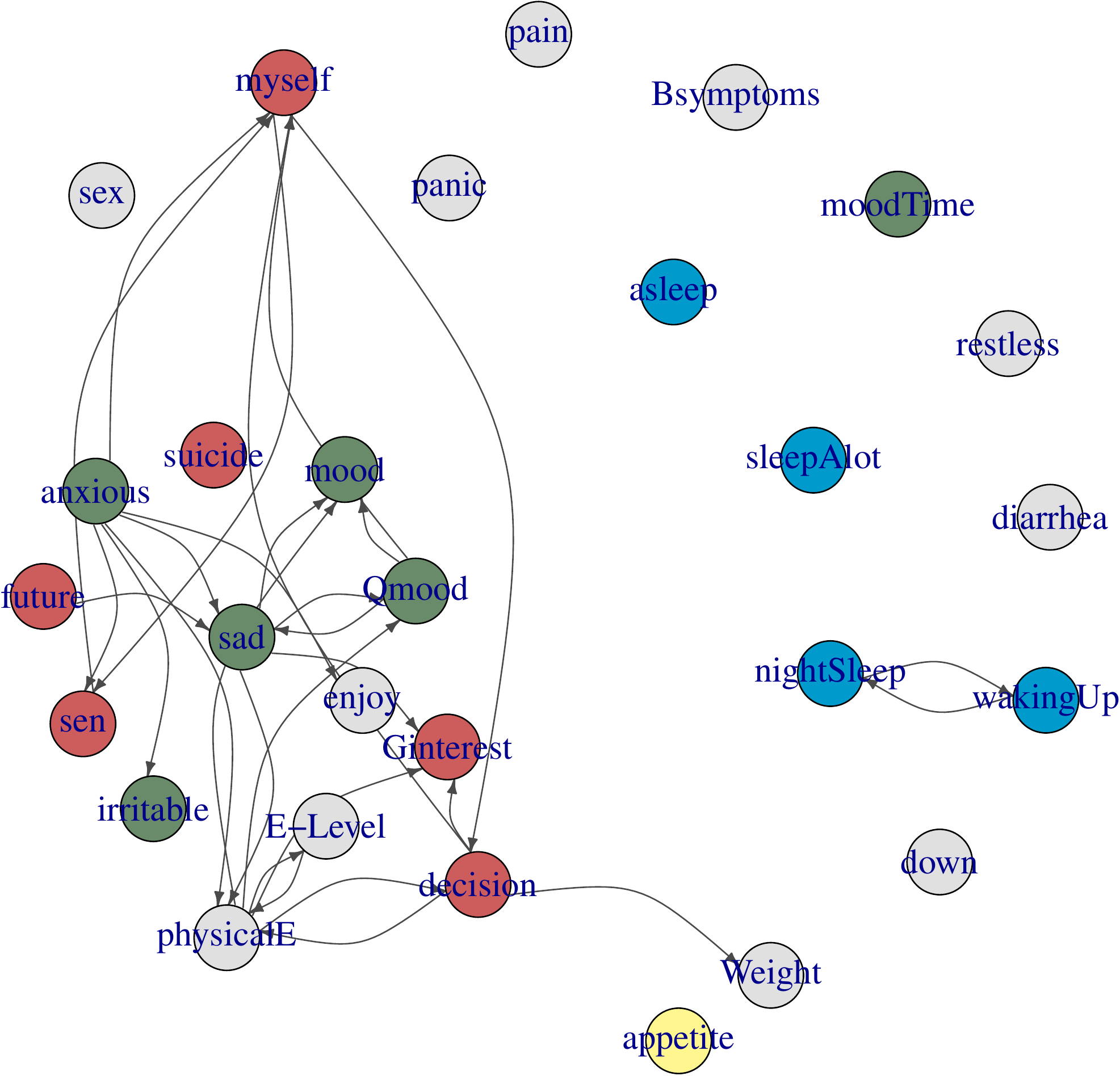}
	
	(a) \hspace{8cm} (b)
	\caption{ Intra time-slice conditional independence undirected network in NESDA dataset (a) and delayed interactions between items in NESDA across time steps(b). There are four categories in NESDA data: (i) sleep in blue, (ii)  mood in green, (iii) appetite in yellow, (iv) somatic in gray, (v) mental in red. }
	\label{nesdaFig}
\end{figure}

First we simulate data from $N_p(0,\Theta^{-1})$ at time $t=1$, for the next time steps $t= 2, \ldots, T$ we use VAR(1) model such that $Z^{(t)} | Z^{(t-1)} \sim N(\Gamma Z^{(t-1)}, \Theta^{-1})$. Then, $n$ i.i.d samples is generated for each time point. This results in p-variate time series data. Finally, we discretize the obtained time series data with Gaussian marginals into randomized quantile ranges and treat them as categorical time series data. The simulations are repeated $50$ times independently for different values of $p$, $n$, $t$.  

To assess the performance of our proposed method in recovering the intra and inter conditional independence relationships we compute the $F_1$-score, sensitivity and specificity measures, which are defined as:
\[
F_1-\mbox{score} = \frac{2TP}{2TP + FP + FN}, \qquad SEN = \frac{TP}{(TP + FN)}, \qquad SPE = \frac{TN}{TN + FP}
\]
where TP, TN, FP, and FN are the numbers of true positive, true negative, false positive, false negative in identifying the non-zero elements in the $\Theta$ and $\Gamma$ matrices. We note that high values of the $F_1$-score, sensitivity and specificity indicate good performance of a method for the given combination of $p$, $n$ and $t$. However, as there is a natural trade off between sensitivity and specificity, we focus particularly on the $F_1$-score to evaluate the performance of each method. 

We compare the finite sample performance of the proposed approach using SCAD penalized maximum likelihood with a recently proposed approach implemented in R package \emph{SparseTSCGM} \citep{abegaz2015package}. For further comparison we have applied \emph{SparseTSCGM} to the original simulated ordinal data and to the transfered data using the normal transformation. We present the simulation results of sparse precision and autoregressive coefficient matrices in Table \ref{tablet5} and Table \ref{tablet10} based on optimal tuning parameters chosen by the minimum EBICs. In each simulation setting, we have very sparse matrices with only $(1/p) \times 100$ nonzero entries. From the tables, we can see that in most cases our method scores better in terms of the F1-score compare with the alternative method. These results suggest that, though recovering sparse network structure in ordinal time series data is a challenging task, the proposed approach has a good performance on model-based simulations. We note here that improved model performance can be gained by allowing the tuning parameters $\rho$ and $\lambda$ to vary with each simulation.  

\section{Netherlands Study of Depression and Anxiety}
\label{application}
We applied our method to a Netherlands Study of Depression and Anxiety (NESDA) Severity of Depression dataset. Depression and anxiety disorders are common at all ages. Approximately one out of three people in the Netherlands will be faced with one of these disorders at some time during their lives. It is still not clear why some people recover quickly and why others suffer for long periods of time. The Netherlands Study of Depression and Anxiety (NESDA) was therefore designed to investigate the course of depression and anxiety disorders over a period of several years. The main aim of NESDA is to determine the (psychological, social, biological and genetic) factors that influence the development and the long-term prognosis of anxiety and depression. The data consist of the $28$ items (variables) that have been collected in $3$ time intervals. For each of $28$ variables there are four corresponding answers 0=None, 1=Mild, 2=Moderate, 3=Severe. For example, for the item “Feeling sad” there are four corresponding answers from “0” that is indicative of no depression (e.g., “I do not feel sad”) to “3” referring to a more severe depressive symptom (e.g., “I feel sad nearly all the time”). A total score is derived (possible range: 0–84), and higher scores are indicative of relatively severe depressive symptomatology. From the 1799 participants, we have selected $200$ patients that have been more informative. The BIC criterion selects the penalty values $\lambda = 0.19$ and $\rho= 0.23$. The resulting instantaneous and delayed interaction network among the $28$ items are shown in Figure \ref{nesdaFig}, left and right panels, respectively. 

Figure \ref{nesdaFig}(a) shows the undirected links that suggest contemporaneous interactions
among 12 items and Figure \ref{nesdaFig}(b) displays the directed edges that indicate granger-causality relationships or delayed interactions between these 12 items. It is observed that item “Feeling sad” is the hub in both figures, suggesting that it plays a fundamental role in treating depression and anxiety disorders. Also, Figure \ref{nesdaFig}(b), shows that there are several directed links pointing from mood category to mental category suggesting that mood disorders influence the development of mental disorders in long term. Interestingly, Figure \ref{nesdaFig}b shows that sleeping disorders do not have any effect on other symptoms of depression.

\section{Discussion}
\label{chap5:Discussion}

We have presented a dynamic model for multivariate ordinal time series data which assumes a chain graph representation of the conditional independence
structure among time series components. The proposed model combines the Gaussian copula graphical models and dynamic Bayesian networks to infer instantaneous conditional dependence relationships among time series components and dynamic or delayed interactions possibly potentially “causal” relationships among variables at consecutive time steps. The directed edges reflect Granger causality whereas the contemporaneous dependence structure is represented by undirected edges. 

To obtain sparse estimates for the instantaneous conditional dependence graph and for the Granger-causality graph, we considered penalized log-likelihood estimation using the $L_1$ and SCAD penalties. Simulation studies show that the proposed sparse estimates reflect the underlying intra- and inter-time slice conditional dependence networks more accurately compared to the only sparse alternative method. 

The method was applied to the Netherlands study of depression and anxiety categorical time series data. The model does, however, have much wider applicability to any multivariate mixed continuous and discrete time series data.  
 
\section{Appendix}
Another approximation that can be replaced in (\ref{firstMoment:inter}) and (\ref{secondMoment:inter}) follows as
\begin{equation}
\label{firstMoment:intra}
E \Big( Z^{(t)}_{i,j}  \Big| y_i; \Theta^\star, \Gamma^\star \Big) = E \Big[ E \Big(Z^{(t)}_{i,j} | Z^{(t)}_{i,-j}, y^{(t)}_{i,j}; \Theta, \Gamma \Big) \ \Big| \ y_i; \Theta^\star, \Gamma^\star \Big]  
\end{equation}
\begin{equation}
\label{secondMoment:intra}
E\Big(Z^{(t)^2}_{i,j} | y_i; \Theta^\star, \Gamma^\star \Big) = E \Big[ E \Big(Z^{(t)^2}_{i,j} | Z^{(t)}_{i,-j}, y^{(t)}_{i,j}; \Theta, \Gamma \Big) \Big| y_i; \Theta^\star, \Gamma^\star \Big]    
\end{equation}
where $Z^{(t)}_{i,-j}$ represents a set that contains all the variables at time step $t$ except the $j$-th variable.

\label{appendix}
In case of within each time step, the mean $\mu_{i,j}$ is a linear function of $z^{(t)}_{i,-j}$, and both $\frac{\phi(\delta_1) - \phi(\delta_2)}{\Phi(\delta_2) - \Phi(\delta_1)}$ and $\frac{\delta_1 \phi(\delta_1) - \delta_2 \phi(\delta_2)}{\Phi(\delta_2) - \Phi(\delta_1)}$ are nonlinear functions of $z^{(t)}_{i,-j}$. Applying Lemma \ref{lemm} on the conditional expectations in (\ref{firstMoment:intra}) and (\ref{secondMoment:intra}) leads to following approximations 
{\footnotesize
\begin{align}
\label{offDiagapp2}
E(Z^{(t)}_{i,j} \ | \ y^{(t)}_i; \Theta^\star, \Gamma^\star) & \approx
 \mathbf{\Sigma}_{j,-j} \mathbf{\Sigma}_{-j,-j}^{-1} E(Z_{i, -j}^{(t)'} \ | \ y^{(t)}_i; \Theta^\star, \Gamma^\star) + \frac{\phi(\delta_{i,j,y_{i,j}^{(t)}}^{(t)}- \phi(\delta_{i,j,y_{i,j}^{(t)}+1}^{(t)})}{\Phi(\delta_{i,j,y_{i,j}^{(t)}+1}^{(t)}) - \Phi(\delta_{i,j,y_{{i,j}}^{(t)}}^{(t)})} \sigma_{j}^{(i)},
\end{align}
\begin{align}
\label{diagapp2}
E((Z^{(t)^2}_{i,j}) \ | \ y^{(t)}_i;  \Theta^\star, \Gamma^\star) &\approx \mathbf{\Sigma}_{j,-j} \mathbf{\Sigma}^{-1}_{-j,-j} E(Z_{i,-j}^{(t)'} Z_{i,-j}^{(t)} \ | \ y^{(t)}_i; \Theta^\star, \Gamma^\star ) \Sigma^{-1}_{-j,-j} \Sigma'_{j,-j} + \sigma^2_{i,j} \nonumber \\
	& + 2 \frac{\phi(\delta_{i,j,y_{i,j}^{(t)}}^{(t)})- \phi(\delta_{i,j,y_{i,j}^{(t)}+1}^{(t)})}{\Phi(\delta_{i,j,y_{i,j}^{(t)}+1}^{(t)}) - \Phi(\delta_{i,j,y_{{i,j}}^{(t)}}^{(t)})} [ \Sigma_{j,-j} \Sigma_{-j,-j}^{-1} E(Z_{i,-j}^{(t)^\tau} \ | \ y^{(t)}_i; \Theta^\star, \Gamma^\star )] \tilde{\sigma}_{i,j}  \nonumber \\
	& + \frac{\delta^{(t)}_{i,j,y_{i,j}^{(t)}} \phi(\delta_{i,j,y_{i,j}^{(t)}}^{(t)}) - \delta_{i,j,y_{i,j}^{(t)}+1}^{(t)} \phi(\delta_{i,j,y_{i,j}^{(t)}+1}^{(t)})}{\Phi(\delta_{i,j,y_{i,j}^{(t)}+1}^{(t)}) - \Phi(\delta_{i,j,y_{i,j}^{(t)}}^{(t)})}\sigma^2_{i,j},
	\end{align}  
}
where $\delta^{(t)}_{i,j,y^{(t)}_{i,j}} = (c^{(t)}_{i,j} - \mu_{i,j}) / \sigma_{ij}$. Here, the first order delta method is used to approximate the nonlinear terms. Moreover, we approximate the elements of conditional expectation matrices $S_{pp}$, $S_{cc}$, and $S_{cp}$ through equations (\ref{offDiagapp2}) and (\ref{diagapp2}).  
\bibliography{Ref5}
\bibliographystyle{Chicago}

\end{document}